\documentclass[12pt,a4paper]{article}%
\usepackage{amssymb}
\usepackage{amsfonts}
\usepackage{amsmath}
\usepackage{graphicx}
\usepackage[singlespacing]{setspace}
\usepackage{geometry}%
\providecommand{\U}[1]{\protect\rule{.1in}{.1in}}
\newtheorem{criterion}{Criterion}

\begin{document}

\title{\textbf{Policy Maker's Credibility with Predetermined Instruments for }\\\textbf{Forward-Looking Targets}\thanks{We thank the editors of \textit{Revue
d'\'{e}conomie politique} and two anonymous referees for their outstanding
comments. PSE thanks the support of the EUR grant ANR-17-EURE-001.}}
\author{Jean-Bernard Chatelain\thanks{Paris School of Economics, Universit\'{e} Paris
I Pantheon Sorbonne, PSE, 48 boulevard Jourdan, 75014 Paris. Email:
jean-bernard.chatelain@univ-paris1.fr} \qquad Kirsten Ralf\thanks{ESCE
International Business School, Inseec U. research center, 10 rue Sextius
Michel, 75015 Paris, Email: Kirsten.Ralf@esce.fr.}}
\maketitle

\begin{abstract}
The aim of the present paper is to provide criteria for a central bank of how
to choose among different monetary-policy rules when caring about a number of
policy targets such as the output gap and expected inflation. Special
attention is given to the question if policy instruments are predetermined or
only forward looking. Using the new-Keynesian Phillips curve with a
cost-push-shock policy-transmission mechanism, the forward-looking case
implies an extreme lack of robustness and of credibility of stabilization
policy. The backward-looking case is such that the simple-rule parameters can
be the solution of Ramsey optimal policy under limited commitment. As a
consequence, we suggest to model explicitly the rational behavior of the
policy maker with Ramsey optimal policy, rather than to use simple rules with
an ambiguous assumption leading to policy advice that is neither robust nor credible.

\textbf{JEL\ classification numbers}: B22, B23, B41, C52, E31, O41, O47.

\textbf{Keywords:} Determinacy, Proportional Feedback Rules, Dynamic
Stochastic General Equilibrium, Ramsey Optimal Policy under Quasi-Commitment.

\textbf{Postprint published in:} \textit{Revue d'\'{e}conomie politique}
(2020), 130(5), 823-846.\bigskip

\textbf{Titre:} Politiques cr\'{e}dibles avec des instruments
pr\'{e}d\'{e}termin\'{e}s pour des cibles non-pr\'{e}d\'{e}termin\'{e}es

\textbf{R\'{e}sum\'{e}:}

Cette artice propose des crit\`{e}res de choix par une banque centrale entre
diff\'{e}rentes r\`{e}gles de politique mon\'{e}taire r\'{e}agissant \`{a}
l'inflation. Une attention particuli\`{e}re est port\'{e}e sur la nature
non-pr\'{e}d\'{e}termin\'{e}e ou pr\'{e}d\'{e}termin\'{e}e des instruments de
politique mon\'{e}taire. En prenant l'exemple d'une courbe de Phillips des
nouveaux Keyn\'{e}siens, un instrument non-pr\'{e}d\'{e}termin\'{e} lorsque
l'inflation est aussi non-pr\'{e}d\'{e}termin\'{e}e implique un manque
extr\^{e}me de cr\'{e}dibilit\'{e} et de robustesse de l'ancrage et de la
stabilisation des prix, d\`{e}s lors que la banque centrale ne conna\^{\i}t
pas exactement les param\`{e}tres de l'\'{e}conomie. En revanche, un
instrument pr\'{e}d\'{e}termin\'{e} est tel qu'une r\'{e}gle proportionnelle
ancre l'inflation non-pr\'{e}d\'{e}termin\'{e}e de mani\`{e}re robuste et
cr\'{e}dible. Cette r\`{e}gle peut alors \^{e}tre une forme r\'{e}duite d'une
politique optimale \`{a} la Ramsey avec un minimum de cr\'{e}dibilit\'{e} et
de robustesse. Nous recommandons de mod\'{e}liser des politique optimales avec
un minimum de cr\'{e}dibilit\'{e} de la banque centrale plut\^{o}t qu'avec des
r\`{e}gles proportionnelles en supposant non-pr\'{e}d\'{e}termin\'{e}s \`{a}
la fois les instruments et les cibles de la politique mon\'{e}taire.

\textbf{Mots cl\'{e}s: }D\'{e}termination, r\`{e}gles de r\'{e}troaction
proportionnelles, Politique de Ramsey avec engagement, mod\`{e}le
d'\'{e}quilibre g\'{e}n\'{e}ral dynamique et stochastique.

\end{abstract}

\section{Introduction}

Throughout the years, a large number of policy rules for the central bank have
been proposed. The aim of the present paper is to provide criteria of how to
choose among these different rules and how to decide in the mathematical
modeling whether the policy instrument should be defined as forward looking or
predetermined and its relation to the policy target. In this context, we will
compare simple rules with a predetermined policy instrument, simple rules with
a forward-looking instrument, and Ramsey optimal policy with quasi commitment
in a model with the new-Keynesian Phillips curve as the transmission mechanism
(Gali (2015). Special importance is given to the problem of robustness and the
existence of bifurcations, following Grandmont (1986), Benhabib,
Schmitt-Groh\'{e} et Uribe (2001a, 2001b, 2003) and Sorger (2005).

In a first part it will be argued that one cannot deduce from a proportional
simple feedback rule, i.e. a rule in which the instrument is a linear function
of the deviations of the policy targets from their long-run equilibrium, that
\emph{the policy instrument is necessarily a forward-looking variable if the
policy target is a forward-looking variable.} An example of a simple rule is
given in Leeper (1991) where the Central Bank's short-term interest rate
responds linearly to inflation deviations. A variable is forward-looking in an
equation not derived from intertemporal optimization, such as a proportional
feedback rule, if two conditions are satisfied: Firstly, the variable should
depend on its own expectations, and not only on the expectations of other
variables. Secondly, these expectations should not be driven by adaptive
expectations. Because these simple proportional rules do not include \emph{the
expectations of the policy instrument itself, }one cannot claim it is obvious
that the policy instrument is a forward-looking variable.

This assumption matters a lot for policy recommendation. Simple rule theorists
solve their model imposing Blanchard and Kahn's (1980) determinacy condition
following an algorithm in three steps (Leeper (1991)):

Firstly, they count the number of predetermined variables.

Secondly, the found number implies the number of stable eigenvalues inside the
unit circle for discrete-time models.

Thirdly, the choice of stable and unstable eigenvalues implies a range of
values of policy-rule parameters for a given policy-transmission mechanism
which is labeled determinacy area. This determinacy area corresponds to policy
advice for policy-rule parameters. This third implication is an outcome of
Wonham's (1967) theorem. For a controllable system, the parameters of the
characteristic polynomial determining these eigenvalues are affine functions
of policy-rule parameters. Eigenvalues within the unit circle correspond to a
bounded interval of values specific to each policy-rule parameter. Conversely,
the values of each policy rule-parameter outside their specific bounded
interval corresponds to eigenvalues outside the unit circle.

What is missing, however, in some of the seminal papers using simple rules
(Leeper (1991), Clarida, Gali and Gertler (1999), Woodford \ (2003), Gali
(2015)), is the discussion of why the policy instrument is a forward-looking
variable instead of a "backward-looking" or "predetermined" variable when its
responds to the deviations of forward-looking policy targets from their set
point. \emph{From their computation of the number of stable eigenvalues}, the
reader can only infer that they \emph{implicitly} assumed that policy
instruments are always forward-looking for simple rules. If policy instruments
are counted as predetermined variables, the number of eigenvalues inside the
unit circle increases. The range of values of policy rule-parameters leading
to determinacy is different. Finally, policy recommendations are different.

The second contribution of this paper is to highlight the consequences of
assuming policy instruments forward-looking instead of predetermined for a
simple rule with forward-looking policy targets. We consider the new-Keynesian
Phillips curve as a monetary policy transmission mechanism (Gali (2015)).
There is an extreme lack of robustness and of credibility if one assumes that
policy instrument is forward-looking when at least one policy target is
forward-looking in a proportional feedback simple rule.

Cochrane (2019, p.345) expressed his concern of the lack of robustness and of credibility:

\begin{quotation}
\emph{To produce those equilibria, the central bank commits that if inflation
gets going, the bank will decrease interest rates, and by doing so it will
increase subsequent inflation, without bound. Likewise, should inflation be
less than the central bank wishes, it will drive the economy down to the
liquidity trap... No central bank on this planet describes its
inflation-control efforts this way. They uniformly explain the opposite.
Should inflation get going, the bank will increase interest rates in order to
reduce subsequent inflation. It will induce stability into an unstable
economy, not the other way around. I have not seen selecting among multiple
equilibria on any central bank's descriptions of what it does... That the
central bank will react to inflation by pushing the economy to hyperinflation
seems an even more tenuous statement about people's beliefs, today and in any
sample period we might study, than it is about actual central bank behavior.}
\end{quotation}

The assumption of forward-looking policy instruments for forward-looking
policy targets in a simple rule corresponds to the complete lack of
credibility of the policy maker. This makes sense: \emph{If a policy maker
does not care at all about the future in his loss function and in the policy
transmission mechanism, he does not care to leave the economy with locally
unstable dynamics of policy targets on future dates. Local unstability will
markedly increases his loss function in the near future if he does not know
the parameters of the policy transmission mechanism with an infinite
precision.} Conversely, policy rules which destabilize the future dynamics of
policy targets are not credible for the private sector.

With some credibility, the initial value of the current policy instrument (the
funds rate) anchors the initial value of the forward-looking policy target
(inflation) using the feedback rule. Hence, there is no longer the
indeterminacy of initial inflation. Therefore, it does not make sense to
destabilize inflation dynamics to rule out indeterminacy. It follows that the
assumption of forward-looking policy instruments for forward-looking policy
targets provides useless policy recommendations for policy makers who are
usually willing to care about the future.

Including \emph{ad-hoc} equations such as a simple proportional feedback rule
for some agents \emph{without leads or lags of the policy instruments}, leads
to ambiguous and misleading results in rational expectations models.
Therefore, introducing an asymmetry between a rational private sector doing
intertemporal optimization and an \emph{ad-hoc} policy rule for policy makers
is not a good idea. Modeling the rational behavior of the policy maker would
have avoided these lacks of robustness and credibility.

Accordingly, assuming the policy instrument to be backward-looking may
correspond to a reduced form of the optimal model of limited commitment, when
the policy maker has a non-zero probability to take into account private
sector's expectations (Roberds (1989), Schaumburg and Tambalotti (2007))).
This maintains an order of stable dynamics (a number of stable eigenvalues
inside the unit circle) which corresponds the order of the dynamics of the
private sector's policy targets. This guarantees the local stability of the
dynamics in the space of the policy targets.

To sum up, the lack of robustness and of credibility of simple rules is an
artefact grounded by a never discussed conventional assumption: \emph{policy
instruments are assumed to be forward-looking when they respond to
forward-looking policy targets in proportional feedback simple rules}.

The rest of the paper proceeds as follows: Section two recalls the
mathematical properties of dynamical models putting a special importance on
the concepts of predetermined versus non-predetermined variables, determinacy
versus indeterminacy, feedback rules, and the convention of defining the
policy instrument as a forward-looking variable. Section three compares the
policy maker's solutions assuming either the policy instrument to be
predetermined or to be backward-looking for a new-Keynesian Phillips curve and
a cost-push shock transmission mechanism. Section four concludes.

\section{Mathematical modeling and economic interpretation}

\subsection{Predetermined versus non-predetermined variables}

One of the first questions when choosing the variables in an economic model is
the question of whether these variable are predetermined or not. Some
variables, such as the stock of capital, seem to be natural candidates for
predetermined variables, for others the decision is more difficult. From a
mathematical point of view, any variable can be predetermined or
non-predetermined. Blanchard and Kahn's (1980) start their analysis of a
dynamical system stating that the matrices $\mathbf{A}$ and $\mathbf{\gamma}$
of their dynamic model are \emph{ad hoc} given matrices, i.e. not derived from
intertemporal optimal choice. The transition matrix $\mathbf{A}$ has a given
number $\overline{n}$ of stable eigenvalues inside the unit circle and a
remaining number $\overline{m}$ of unstable eigenvalues outside the unit
circle. They define predetermined (backward-looking) variables and
non-predetermined (forward-looking) variables and their model as follows:

\begin{quotation}
``The model is given as follows:%
\begin{equation}
\left(
\begin{array}
[c]{c}%
\mathbf{X}_{t+1}\\
_{t}\mathbf{P}_{t+1}%
\end{array}
\right)  =\mathbf{A}\left(
\begin{array}
[c]{c}%
\mathbf{X}_{t}\\
\mathbf{P}_{t}%
\end{array}
\right)  +\mathbf{\gamma Z}_{t},\qquad\mathbf{X}_{t=0}=\mathbf{X}_{0}%
\end{equation}
where $\mathbf{X}$ is an $\left(  n\times1\right)  $ vector of variables
predetermined at $t$; $\mathbf{P}$ is an $\left(  m\times1\right)  $ vector of
variables non-predetermined at $t$; $\mathbf{Z}$ is an $\left(  k\times
1\right)  $ vector of exogenous variables; $_{t}\mathbf{P}_{t+1}$ is the
agents expectations of $\mathbf{P}_{t+1}$ held at $t$; $\mathbf{A}%
$,$\mathbf{\gamma}$ are $\left(  n+m\right)  \times\left(  n+m\right)  $ and a
$\left(  n+m\right)  \times k$ matrices, respectively.%
\begin{equation}
_{t}\mathbf{P}_{t+1}=E_{t}\left(  \mathbf{P}_{t+1}\shortmid\Omega_{t}\right)
.
\end{equation}
where $E_{t}\left(  \cdot\right)  $ is the mathematical expectation operator;
$\Omega_{t}$ is the information set at date $t$,.... A \textbf{predetermined}
variable is a function only of variables known at date $t$, that is of
variables in $\Omega_{t}$ that $\mathbf{X}_{t+1}=$ $_{t}\mathbf{X}_{t+1}$
whatever the realization of the variables in $\Omega_{t+1}$. A
\textbf{non-predetermined} variable can be a function of any variable in
$\Omega_{t+1}$, so that we can conclude that $\mathbf{P}_{t+1}=$
$_{t}\mathbf{P}_{t+1}$ only if the realization of all variables in
$\Omega_{t+1}$ are equal to their expectations conditional on $\Omega_{t}$.''
Blanchard and Kahn (1980), p.1305.
\end{quotation}

Blanchard and Kahn (1980) thus assume that the property of whether a variable
is predetermined or forward-looking and their respective numbers $n$ and $m$
is given. In practice, they allow that $n$, $m$, $\overline{n}$, $\overline
{m}$ are discretionary \emph{ad hoc} choices done by the researcher as it is
done by other \emph{ad hoc} linear rational expectations models of the 1970s.

As an example they give an \emph{ad hoc} multiplier accelerator model
including only one non-predetermined (or forward-looking or jump) variable:
output (example 4). They show that a researcher can choose which economic
variable is a predetermined variable (assuming an equation for the dynamics of
this variable corresponding to the first line of the system (1)) or a
non-predetermined variable (assuming an equation for the dynamics of this
variable corresponding to the second line of the system (1)). This choice is
independent from an economic criterion such as real versus nominal variables (p.1307):

\begin{quotation}
This example also shows the absence of necessary connection between "real,"
"nominal" and "predetermined," "non-predetermined".
\end{quotation}

Set aside this sentence, \emph{Blanchard and Kahn's (1980) paper does not
provide any economic or mathematical criterion for deciding when a variable is
predetermined (with dynamics within the first equation of system (1)) or
non-predetermined (with dynamics within the second equation of system (1)).
}In their example of the multiplier accelerator model, the demand components
of output: consumption, investment and government expenditures \emph{may all
have been assumed} to depend on forward-looking specific expectations of
consumption, investment and government expenditures, instead of depending only
on aggregate output expectations.

\subsection{Optimal control and feedback rule}

The next choice of the researcher when modeling monetary policy concerns the
type of feedback rule of the central bank. Currie and Levine (1985), Leeper
(1991), Clarida, Gali and Gertler (1999), Woodford (2003) and Gali (2015) add
to Blanchard and Kahn's (1980) dynamic system at least one policy instrument
$i_{t}$ with non-zero correlation ($\mathbf{B\neq0}$) with policy targets
$\left(  \mathbf{X}_{t},\mathbf{P}_{t}\right)  $:%

\begin{equation}
\left(
\begin{array}
[c]{c}%
\mathbf{X}_{t+1}\\
_{t}\mathbf{P}_{t+1}%
\end{array}
\right)  =\mathbf{A}\left(
\begin{array}
[c]{c}%
\mathbf{X}_{t}\\
\mathbf{P}_{t}%
\end{array}
\right)  +\mathbf{\gamma Z}_{t}+\mathbf{B}i_{t},\qquad\mathbf{X}%
_{t=0}=\mathbf{X}_{0}.
\end{equation}

In their examples, Kalman controllability, i.e. the property that policy
instruments have a direct of indirect effect on the policy target, is satisfied:%

\[
rank\left(  \mathbf{B,AB,A}^{2}\mathbf{B,...A}^{n+m-1}\mathbf{B}\right)
=n+m.
\]

The policy maker's rule is assume to be an ad hoc (behavioral) proportional
feedback rule or "simple rule" with policy-rule parameters $\mathbf{F}_{X}$,
$\mathbf{F}_{P}$ and $\mathbf{F}_{X}$:%

\begin{equation}
i_{t}=\mathbf{F}_{X}\mathbf{X}_{t}+\mathbf{F}_{P}\mathbf{P}_{t}+\mathbf{F}%
_{Z}\mathbf{Z}_{t} . \label{rule}%
\end{equation}

Restrictions on the policy-rule parameters are often imposed. For example,
Leeper's (1991) interest-rate rule responds only to forward-looking inflation
and to an exogenous auto-regressive variable ($\mathbf{F}_{X}=\mathbf{0}$).

These proportional feedback rules exhibit \emph{no dynamics}:

- The policy instrument $i_{t}$ does not depend on its lagged values
$i_{t-1},$ $i_{t-2}..$. It is not included in the set of predetermined
variables in the first row of Blanchard and Kahn (1980) system: $\mathbf{X}%
_{t+1}=\mathbf{A}_{nn}\mathbf{X}_{t}+...$ where $\mathbf{A}_{nn}$ is the upper
left $n\times n$ square block matrix in matrix of matrix $\mathbf{A}$.

- \emph{The policy instrument }$i_{t}$\emph{ does not depends on its expected
value:} $_{t}i_{t+1}=E_{t}\left(  i_{t+1}\right)  $. It is not included in the
set of non-predetermined (forward-looking) variables in the second row of
Blanchard and Kahn (1980) system: $_{t}\mathbf{P}_{t+1}=\mathbf{A}%
_{mm}\mathbf{P}_{t}+...$ where $\mathbf{A}_{mm}$ is the bottom right $m\times
m$ square (block matrix of matrix $\mathbf{A}$.

The following choice of a policy rule, however, would allow the policy
instrument to be forward looking or backward looking:%

\begin{equation}
i_{t}=\mathbf{F}_{X}\mathbf{X}_{t}+\mathbf{F}_{P}\mathbf{P}_{t}+\mathbf{F}%
_{Z}\mathbf{Z}_{t}+\beta_{1}i_{t-1}+\beta_{2}i_{t-2}+...+b_{1}E_{t}\left(
i_{t+1}\right)  .
\end{equation}

If the policy instrument depends on its lagged value $\beta_{i}\neq0$ and not
on its expectation $b_{1}=0$, then $i_{t}$ is a backward-looking variable. If
the variable does not depend on its lagged value $\beta_{i}=0$, but it depends
on its expected value $b_{1}\neq0$ and the expectations are not adaptive
expectations, i.e. its expected value does not depend on the current and
lagged values of the policy instrument so that they "\emph{cause}" the current
value $i_{t}$, then $i_{t}$ is a forward-looking variable.

In the hybrid case, where the policy instrument does depend on its lagged
value $\beta_{i}\neq0$ and also on its expected value $b_{1}\neq0$, the policy
instrument $i_{t}$ is still assumed to be a non-predetermined variable. With
one lag $\beta_{1}\neq0$ and one expected value $b_{1}\neq0$, the dynamics of
the policy instrument is of order two. Its solution involves one unstable
eigenvalue and one stable eigenvalue (Chatelain and Ralf (2018)).

This paper focuses only on the following case. If the feedback rule has no
dynamics such that $\beta_{i}=0$ and $b_{1}=0$, the above cited definitions of
predetermined versus non-predetermined variables of Blanchard and Kahn (1980)
do not imply that the policy instrument is forward-looking when at least one
policy target is forward-looking.

\subsection{Determinacy versus indeterminacy}

An important feature of a rational expectations equilibrium in a dynamic model
is the indeterminacy or determinacy of the perfect foresight dynamics.
Determinacy requires the steady state equilibrium to be locally unique,
whereas in the case of indeterminacy a large number of nonstationary solutions
of the dynamical system converging to the steady state exist. Then it is for
the individual of no importance which future state of nature he expects
because for any choice of today's actions the economy approaches the steady
state, no rational decision is possible.

To avoid this kind of ambiguity, Blanchard and Kahn (1980) assumed a stability
boundary condition in the infinite horizon, by analogy with the infinite
horizon transversality optimality condition in optimal control. Because
Blanchard and Kahn (1980) model (1) is not necessarily related to
intertemporal maximization, their stability boundary condition in the infinite
horizon is only an \emph{assumption}. By contrast, in optimal control, the
transversality condition is an \emph{optimality condition}.

Then, Blanchard and Kahn (1980) derive conditions for determinacy,
indeterminacy and for a bounded solution:

\begin{quotation}
``This condition in effect rules out exponential growth of the expectations of
$\mathbf{X}_{t+1}$ and $\mathbf{P}_{t+1}$ held at time $t$. (This in
particular rules out "bubbles"...).'' Blanchard and Kahn (1980), p.1307.

``Proposition 1: If $m=\overline{m}$, i.e., if the number of eigenvalues of
$\mathbf{A}$ outside the unit circle is equal to the number of
non-predetermined variables, then there exists a unique solution.'' Blanchard
and Kahn (1980), p.1308.

``Proposition 2: If $\overline{m}<m$, i.e., if the number of eigenvalues of
$\mathbf{A}$ outside the unit circle exceeds the number of non-predetermined
variables, there is solution satisfying both (1) and the non-explosion
condition.'' Blanchard and Kahn (1980), p.1308.

``Proposition 3: If $\overline{m}>m$, i.e., if the number of eigenvalues of
$\mathbf{A}$ outside the unit circle is less than the number of
non-predetermined variables, there is an infinity of solutions.'' Blanchard
and Kahn (1980), p.1308.
\end{quotation}

Optimal intertemporal choice models always satisfy the determinacy condition
under adequate concavity and boundary conditions: this has been proven in
papers of optimal control before Blanchard and Kahn (1980), such as Vaughan
(1970) and Kalman (1960). Blanchard and Kahn's (1980) solution is an extension
to \emph{ad hoc} models (not derived from maximization) of Vaughan (1970)
solution of the maximization of a quadratic function subject to linear
constraints (linear quadratic regulator, LQR).

\begin{quotation}
``How likely are we to have $m=\overline{m}$ in a particular model? It may be
that the system described by (1) is just the set of necessary conditions for
maximization of a quadratic function subject to linear constraints. In this
case, the matrix $\mathbf{A}$ will have more structure than we have imposed
here, and the condition $m=\overline{m}$ will always hold.'' Blanchard and
Kahn (1980), p.1309.
\end{quotation}

The matrix $\mathbf{A}_{2n}$ of the Hamiltonian system of the LQR is
symplectic (its transpose is equivalent to its inverse), which implies
$\overline{n}=\overline{m}$. Its number $\overline{n}$ of eigenvalues inside
is the unit circle is equal to its number $\overline{m}$ of eigenvalues
outside the unit circle. This number is equal to its number $n$ of
predetermined state variables and its number $m=n$ of non-predetermined
costate variables.

The linear quadratic model is also a local approximation around an equilibrium
of non-quadratic non-linear optimal control problems such as optimal saving
and optimal growth models under adequate concavity assumptions. These optimal
saving and optimal growth models can be solved by the Lagrange method (but not
necessarily) with Hamiltonian systems having the same saddle-point property as
the linear quadratic regulator:

\begin{quotation}
``The condition $m=\overline{m}$ is also clearly related to the strict saddle
point property discussed in the context of growth models: the above
proposition (1) states in effect that a unique solution will exist if and only
if $\mathbf{A}$ has the strict saddle point property.'' Blanchard and Kahn
(1980), p.1309.
\end{quotation}

Since Blanchard and Kahn (1980) has been published, the macroeconomic theory
of dynamic stochastic general equilibrium (DSGE) models shifted to models
based on intertemporal optimization of the private sector agents
\emph{linearized} around an equilibrium. Hence, Blanchard and Kahn's (1980)
determinacy condition is \emph{always fulfilled} for the private sector part
of the new-Keynesian DSGE\ models or models of the fiscal theory of the price level.

Blanchard and Kahn's (1980) determinacy condition is only a contribution which
matters for \emph{ad hoc} linear models including forward-looking variables.
In linearized DSGE\ models, Blanchard and Kahn's (1980) indeterminacy
condition only matters because of the assumption of policy maker's \emph{ad
hoc} proportional feedback rule ("simple rules") such as a Taylor or
funds-rate rule.

\subsection{The Convention of a Policy Instrument as a Forward-Looking
Variable: Leeper (1991)}

Leeper's (1991) is one of the first papers where the private sector follows an
explicit optimal intertemporal behaviour and where the policy maker behaviour
is described by proportional feedback rule. This seminal paper is an example
of how was set the convention that a policy instrument is \emph{a
forward-looking variable} \emph{when at least one of its policy target is
forward-looking} in the simple proportional feedback rule.

Leeper (1991) considers an intertemporal frictionless consumption model where
the representative household receives every period a constant endowment of
consumption goods and saves in public debt or in money. There is no
production. Lump-sum taxes and nominal rates of return on bonds can vary over time.

Public debt is a private sector state variable. The accounting stock flow
equation for public debt does not involve expected values. Public debt can be
deduced to be a predetermined variable even though Leeper does not state this
explicitly nor gives its initial condition.

Consumption is the private sector optimal decision variable. It is related to
a costate variable. Because the endowment of consumption goods is the same for
every period, consumption is also the same in all periods. The Euler
conditions boils down to a Fisher equation. The nominal return on bonds less
expected inflation is equal to the time-invariant household's real discount rate.

Final boundary (transversality) conditions are assumed for the stocks of real
debt and real balances, but not for inflation. The model is frictionless in
the sense that prices are not set by households. Inflation expectations are
described as conditional expectation taken with respect to an information set
containing all current and past individual variables. Hence, we deduce
inflation is a forward-looking variable.

Leeper introduces an ad hoc interest rate rule where the monetary authority
sets the current nominal rate as a function of current inflation (which is a
forward-looking variable) and an auto-regressive shock with additive normal
disturbances (which is a backward-looking variable).

He introduces an ad hoc fiscal rule where the fiscal authority adjusts direct
lump-sum taxes in response to the level of real government debt outstanding
(which is a backward-looking variable) and to an auto-regressive shock (which
is a backward-looking variable).

He substitutes the nominal rate by the interest rate rule in the Fisher
equation. Expected inflation is correlated with current inflation, after
elimination of the nominal rate. He substitutes lump-sum tax in the stock-flow
accounting equation for public debt. He then has two dynamic equations for
predetermined public debt and for forward-looking inflation where the policy
instruments (nominal funds rate and lump-sump taxes) have been eliminated. He
states "\emph{A sufficient condition for a unique saddle-path equilibrium is
that one root of the system lies inside the unit circle and one root lies
outside [Blanchard and Kahn (1980)]}". At this stage only, \emph{we deduce}
that the nominal rates and lump-sum taxes have been assumed to be
forward-looking, \emph{without any explanation}.

This assumption and the determinacy condition implies one root of the system
lying inside the unit circle and one root outside the unit circle instead of
two roots of the system lying inside the unit circle. Leeper's (1991)
determinacy solution implies the local instability of the policy targets
dynamics in the space of dimension two of the two policy targets (inflation
and public debt).

Finally, the stock of money is a function of the current nominal return
without leads or lags of money, hence, without dynamics. Later in the paper it
is mentioned to have an initial value. Hence, it is a predetermined variable,
by contrast to policy instruments.

\subsection{The Analogy of Game against Nature versus Dynamic Stackelberg game
in Leeper (1991)}

Blanchard and Kahn (1980) is a transpose by analogy to ad hoc models of
Vaughan (1970) optimal linear quadratic regulator solution, which corresponds
to a game against nature. A single agent is controlling the local stability of
predetermined state variables given by nature, using forward-looking co-state variables.

We can equally transpose by analogy the Stackelberg dynamic game optimal
solution also mentioned as Ramsey optimal policy since Kydland and Prescott
(1980). The policy maker is a leader minimizing a loss function which takes
into account state variables (from the point of view of the policy maker)
which include private sector's predetermined state variables \emph{and}
private sector's forward-looking (co-state) variables. The policy maker will
then have its own forward-looking co-state variables for the private sector's
predetermined state variables.

\emph{The policy maker's co-state variable of a private sector co-state
variable} corresponds to the marginal value of the policy maker's loss
function with respect to this private sector co-state variable (for example,
forward-looking inflation). The policy maker first-order condition implies
that \emph{the policy maker's co-state variable of a private sector
forward-looking variable is predetermined at zero at the initial date}. This
optimal initial transversality condition provides the initial value of the
policy maker's instrument and of the private sector forward-looking variable
(for example inflation). As a Stackelberg leader, the policy maker decides (or
anchors) optimally the initial value of private sector's jump variables, using
the initial value of his policy instrument, if he has at least a minimal credibility.

Transposing Stackelberg dynamic game to a proportional feedback rule where the
policy instrument responds to at least one forward-looking private sector
variable, the analogy suggests that the policy instrument is a predetermined variable.

In Leeper's model, the nominal rate is controlling the dynamics of
forward-looking inflation. From the proportional feedback rule equation
\emph{without leads or lags of nominal rates}, \emph{nothing forbids to assume
that the nominal rate is a predetermined variable}. Then, the initial
condition for inflation is found inverting the rule: $\pi_{0}=F^{-1}%
i_{0}+F^{-1}z_{0}$, where $z_{0}$ is the initial value of the monetary policy
shock and $F$ the policy rule parameter.

By contrast, lump-sum tax is controlling the dynamics of predetermined public
debt. The initial value of the lump-sum tax is then a forward-looking variable
where the initial value is given by the initial public debt and the initial
value of the fiscal shock: $\tau_{0}=Gb_{0}+u_{0}$ where $z_{0}$ is the
initial value of the monetary policy shock and $G$ the fiscal rule parameter.

If one follows the analogy with a Stackelberg dynamic game, there are two
predetermined variables: the private sector's public debt and the policy
maker's funds rate. Two roots of the system lie inside the unit circle and
satisfy Blanchard and Kahn's (1980) determinacy condition. The dynamic system
implies locally stable dynamics in the space of the two policy target
variables which has two dimensions, see Chatelain and Ralf (2020a) solution of
Leeper's (1991) policy transmission mechanism.

\section{Example: New-Keynesian Phillips curve}

In order to clarify the possible choice of a policy instrument, we will
present different scenarios of forward-looking or predetermined instruments in
a model where the transmission mechanism is based on the New-Keynesian
Phillips curve.

Some researchers may appeal to\emph{ a criterion based on an analogy with
another underlying model} in order to ground that policy instruments are
forward-looking or predetermined. For example, the criterion: "The central
bank can choose its funds rate without any reference to the past" corresponds
to the underlying model of discretion in Gali (2015), chapter 5.

Conversely, the case for a predetermined policy instrument so that "The
central bank can choose its funds rate with some reference to the past"
corresponds to Ramsey optimal policy under quasi-commitment (Schaumburg and
Tambalotti (2007)) where the probability of not reneging commitment can be
very close to zero, but not exactly zero as it is the case in the discretion model.

\subsection{Policy transmission mechanism}

Following Gali (2015), the monetary policy transmission mechanism is derived
from the new-Keynesian Phillips curve:%

\begin{equation}
\pi_{t}=\beta E_{t}\pi_{t+1} +\kappa x_{t}+z_{t}\text{ where }\kappa>0\text{,
}0<\beta<1\text{, }%
\end{equation}

where $x_{t}$ represents the output gap, i.e. the deviation between (log)
output and its efficient level. $\pi_{t}$ denotes the rate of inflation
between periods $t-1$ and $t$. $\beta$ denotes the discount factor. $E_{t}$
denotes the expectation operator. The cost push shock $z_{t}$ includes an
exogenous auto-regressive component:%

\begin{equation}
z_{t}=\rho z_{t-1}+\varepsilon_{t}\text{ where }0<\rho<1\text{ and
}\varepsilon_{t}\text{ i.i.d. normal }N\left(  0,\sigma_{\varepsilon}%
^{2}\right)  ,
\end{equation}

where $\rho$ denotes the auto-correlation parameter and $\varepsilon_{t}$ an
exogenous shock, identically and independently distributed (i.i.d.) following
a normal distribution with constant variance $\sigma_{\varepsilon}^{2}$. The
transmission mechanism can also include an Euler consumption equation
(Chatelain and Ralf (2020b)) and a public debt equation (Chatelain and Ralf (2020c)).

The transmission mechanism can be written:%

\begin{equation}
\left(
\begin{array}
[c]{c}%
E_{t}\pi_{t+1}\\
z_{t+1}%
\end{array}
\right)  =\left(
\begin{array}
[c]{cc}%
\mathbf{A}_{yy}=\frac{1}{\beta} & \mathbf{A}_{yz}=-\frac{1}{\beta}\\
\mathbf{A}_{zy}=0 & \mathbf{A}_{zz}=\rho
\end{array}
\right)  \left(
\begin{array}
[c]{c}%
\pi_{t}\\
z_{t}%
\end{array}
\right)  +\left(
\begin{array}
[c]{c}%
\mathbf{B}_{y}=-\frac{\kappa}{\beta}\\
\mathbf{B}_{z}=0
\end{array}
\right)  x_{t}+\left(
\begin{array}
[c]{c}%
0\\
1
\end{array}
\right)  \varepsilon_{t+1}.
\end{equation}

The system is already written in the Kalman canonical form which is defined
such that the bottom left block matrix of $\mathbf{A}$ is equal to zero
$\mathbf{A}_{zy}=0$ and the bottom block matrix of $\mathbf{B}$ is zero:
$\mathbf{B}_{z}=0$. It does not, however, satisfy the Kalman controllability condition:%

\begin{align*}
\mathbf{AB}  &  =\left(
\begin{array}
[c]{cc}%
\frac{1}{\beta} & -\frac{1}{\beta}\\
0 & \rho
\end{array}
\right)  \left(
\begin{array}
[c]{c}%
-\frac{\kappa}{\beta}\\
0
\end{array}
\right)  =\left(
\begin{array}
[c]{c}%
-\frac{\kappa}{\beta^{2}}\\
0
\end{array}
\right) \\
rank\left(  \mathbf{B|AB}\right)   &  =rank\left(
\begin{array}
[c]{cc}%
-\frac{\kappa}{\beta} & -\frac{\kappa}{\beta^{2}}\\
0 & 0
\end{array}
\right)  =1<n=2.
\end{align*}

To make the system controllable two additional assumptions can be made.

\textbf{Assumption 1}: Considering only the first equation, there is one
policy target, $n=1$, which leads to the Kalman controllability condition:
$rank\left(  \mathbf{B}_{y}\right)  =rank\left(  -\frac{\kappa}{\beta}\right)
=n=1$, if $-\frac{\kappa}{\beta}\neq0$. If we assume that the slope of the
new-Keynesian Phillips curve is not equal to zero: $\kappa\neq0$, then
\emph{inflation is controllable by the output gap}. There is a non-zero
correlation between the policy instrument (output gap $u_{t}$) and the
expected value of the policy target (inflation $E_{t}\pi_{t+1}$).

\textbf{Assumption 2}: Considering only the second equation, the Kalman
controllability condition is: $rank\left(  \mathbf{B}_{z}\right)  =rank\left(
0\right)  =0<1$. The cost-push shock is not controllable by the output gap.
There is a zero correlation between the policy instrument (output gap $u_{t}$)
and the future value of the cost-push forcing variable $z_{t+1}$. If we assume
that the non-controllable cost-push forcing variable is stationary,
$0<A_{zz}=\rho<1$, then \emph{the dynamic system including both equations can
be stabilized}.

\subsection{Simple rule with a predetermined policy instrument}

Clarida, Gali and Gertler (1999) and Gali (2015, chapter 5) label a rule where
the policy instrument is the output gap $x_{t}$ and the policy target is
forward-looking inflation a "targeting" rule. The policy maker follows a
feedback rule where the output gap does not depend on its expectations nor on
its lagged values:%

\[
x_{t}=F_{\pi}\pi_{t}+F_{z}z_{t}.
\]

After substitution of the policy rule, we get the following "closed loop"
linear system:%

\[
\left(
\begin{array}
[c]{c}%
E_{t}\pi_{t+1}\\
z_{t+1}%
\end{array}
\right)  =\left(
\begin{array}
[c]{cc}%
\lambda_{SR}=\frac{1-\kappa F_{\pi}}{\beta} & \frac{-1-\kappa F_{z}}{\beta}\\
0 & \rho
\end{array}
\right)  \left(
\begin{array}
[c]{c}%
\pi_{t}\\
z_{t}%
\end{array}
\right)  +\left(
\begin{array}
[c]{c}%
0\\
1
\end{array}
\right)  \varepsilon_{t+1}.
\]

When the policy instrument is pegged at its long run optimal value $x_{t}=0$
at all dates ($F_{\pi}=$ $F_{z}=0$), the dynamics is defined as "open-loop
dynamics". Because $\frac{1}{\beta}>1$, the open-loop dynamics of inflation is unstable.

Let us assume that the policy instrument is predetermined with a given initial
value of the policy instrument $x_{0}$ and of the auto-regressive cost-push
shock $z_{0}$. Blanchard and Kahn's (1980) determinacy condition implies two
stable eigenvalues. As $0<\rho<1$, this implies $\left\vert \frac{1-\kappa
F_{\pi}}{\beta}\right\vert <1$. As a definition, "negative feedback" values of
the policy rule parameter $F_{\pi}$ are such that the closed-loop system has
stable dynamics. Blanchard and Kahn's (1980) determinacy condition implies the
following "determinacy set" $D_{NF}$ for the negative-feedback values of the
policy-rule parameter for given parameters of the transmission mechanism
$\left(  \beta,\kappa\right)  $:%

\[
F_{\pi}\in D_{NF}=\left\{  F_{\pi}\in%
%TCIMACRO{\U{211d} }%
%BeginExpansion
\mathbb{R}
%EndExpansion
\text{ such that }\left\vert \frac{1-\kappa F_{\pi}}{\beta}\right\vert
<1\text{ }\right\}  =\left]  \frac{1-\beta}{\kappa},\frac{1+\beta}{\kappa
}\right[  .
\]

Finally, the policy rule allows to find the initial value of the
forward-looking variable.%

\[
x_{0}=F_{\pi}\pi_{0}+F_{z}z_{0}\Rightarrow\pi_{0}=F_{\pi}^{-1}x_{0}-F_{\pi
}^{-1}F_{z}z_{0}\text{.}%
\]

The dynamic system is given in table 1 on the row where $x_{0}$ is known.

\subsection{Ramsey optimal policy with quasi commitment}

Schaumburg and Tambalotti (2007) introduce a sequence of policy regimes
indexed by $j$, in which a policy maker $j$ may re-optimize, i.e. minimize
again its loss function, on each future period with exogenous probability
$1-q$ strictly below one. The length of their tenure or \textquotedblleft
regime" depends on a sequence of exogenous i.i.d. Bernoulli signals $\left\{
\eta_{t}\right\}  _{t\geq0}$ with $E_{t}\left[  \eta_{t}\right]  _{t\geq
0}=1-q$, with $0<q<1$. If $\eta_{t}=1,$ a new policy maker indexed by $k$
takes office at the beginning of time $t$, otherwise the policy maker $j$
stays on. A higher probability $q$ can be interpreted as a higher credibility.
Starting from regime $j$ till a regime change to $k$, the policy maker
maximizes the following function:%

\begin{align}
V^{jk}\left(  u_{0}\right)   &  =-E_{0}%
%TCIMACRO{\dsum \limits_{t=0}^{t=+\infty}}%
%BeginExpansion
{\displaystyle\sum\limits_{t=0}^{t=+\infty}}
%EndExpansion
\left(  \beta q\right)  ^{t}\left[  \frac{1}{2}\left(  \pi_{t}^{2}%
+\frac{\kappa}{\varepsilon}x_{t}^{2}\right)  +\beta\left(  1-q\right)
V^{jk}\left(  u_{t}\right)  \right]  ,\\
\text{s.t. }\pi_{t}  &  =\kappa x_{t}+\beta qE_{t}\pi_{t+1}+\beta\left(
1-q\right)  E_{t}\pi_{t+1}^{k}+z_{t}\text{ (Lagrange multiplier }\gamma
_{t+1}\text{)}\nonumber\\
z_{t}  &  =\rho z_{t-1}+\eta_{t}\text{,}\forall t\in%
%TCIMACRO{\U{2115} }%
%BeginExpansion
\mathbb{N}
%EndExpansion
\text{, }z_{0}\text{ given.}\nonumber
\end{align}

The loss function corresponds to households welfare with $\alpha=\frac{\kappa
}{\varepsilon}>0$, where $\varepsilon=\frac{\kappa}{\alpha}>1$ is the
elasticity of substitution between differentiated goods (Gali (2015)). The
policy target is inflation and the policy instrument is the output gap (Gali
(2015)). The utility of the central bank if next period objectives change is
denoted $V^{jk}$. \ Inflation expectations depend on two terms. The first term
with weight $q$, is the inflation that would prevail under the current regime
$j$ to which the central bank has made a commitment to. The second term with
weight $1-q$, is the inflation that would be implemented under the alternative
regime $k$, which is taken as exogenous by the current central bank. The key
change is that the narrow range of values for the discount factor usually
around $0.99$ for quarterly data ($4\%$ discount rate) is much wider for the
"credibility weighted discount factor" of the policy maker: $\beta q\in\left]
0,0.99\right]  $. A policy maker with lower credibility gives more weight to
the present and near future welfare losses.

Differentiating the Lagrangian with respect to the policy instrument (output
gap $x_{t}$) and to the policy target (inflation $\pi_{t}$) yields the first
order conditions:%

\[
\left\{
\begin{array}
[c]{c}%
\frac{\partial L}{\partial\pi_{t}}=0:\pi_{t}+\gamma_{t+1}-\gamma_{t}=0\\
\frac{\partial L}{\partial x_{t}}=0:\frac{\kappa}{\varepsilon}x_{t}%
-\kappa\gamma_{t+1}=0
\end{array}
\right.  \Rightarrow\left\{
\begin{array}
[c]{c}%
x_{t}=x_{t-1}-\varepsilon\pi_{t}\\
x_{t}=\varepsilon\gamma_{t+1}=\varepsilon(\gamma_{t}-\pi_{t})
\end{array}
\right.
\]

$t=1,2,...$ The central bank's Euler equation ($\frac{\partial L}{\partial
\pi_{t}}=0$) links recursively the future or current value of central bank's
policy instrument $x_{t}$ to its current or past value $x_{t-1}$, because the
central bank's relative cost of changing her policy instrument is strictly
positive $\alpha_{x}=\frac{\kappa}{\varepsilon}>0$. This non-stationary Euler
equation adds an unstable eigenvalue when analyzing the central bank's
Hamiltonian system including three laws of motion of one forward-looking
variable (inflation $\pi_{t}$) and of two predetermined variables $\left(
z_{t},x_{t}\right)  $ or $\left(  z_{t},\gamma_{t}\right)  $.

Using Chatelain and Ralf's (2019) algorithm, the root of the dynamic system,
that we denote "\emph{inflation eigenvalue}" $\lambda$, is inside an interval
belonging to the unit circle for $\varepsilon\in\left]  1,+\infty\right[  $:
\[
\lambda=\frac{1-\kappa F_{\pi}}{\beta q}=\frac{1}{2}\left(  1+\frac{1}{\beta
q}+\frac{\varepsilon\kappa}{\beta q}\right)  -\sqrt{\frac{1}{4}\left(
1+\frac{1}{\beta q}+\frac{\varepsilon\kappa}{\beta q}\right)  ^{2}-\frac
{1}{\beta q}}\in\left]  0,\frac{1}{\beta q}\right[  .
\]

The optimal policy-rule parameters of a linear quadratic optimal program do
not depend on the initial conditions $\left(  x_{0},\pi_{0},z_{0}\right)  $
(Simon (1956)) but only on policy transmission mechanism parameters $\left(
\beta,\kappa\right)  $ and on the policy maker's cost of changing the policy
instrument ($\alpha>0$):
\[
F_{\pi}^{\ast}=\frac{1-\beta q\lambda}{\kappa}=\varepsilon\frac{\lambda
}{1-\lambda}\text{ and }F_{z}^{\ast}=\frac{-1}{1-\beta q\rho\lambda}F_{\pi
}^{\ast}.
\]

This leads to a smaller set $D_{NF}^{\ast}\subset D_{NF}$ when varying the
policy maker's preferences for $\varepsilon\in\left]  1,+\infty\right[  $:%

\[
F_{\pi}^{\ast}\in D_{NF}^{\ast}=\left\{  F_{\pi}^{\ast}\left(  \varepsilon
\right)  \in%
%TCIMACRO{\U{211d} }%
%BeginExpansion
\mathbb{R}
%EndExpansion
\text{ such that }0<\frac{1-\kappa F_{\pi}^{\ast}\left(  \varepsilon\right)
}{\beta q}<\frac{1}{\beta q}\text{ for }\varepsilon\in\left]  1,+\infty
\right[  \right\}  \subset D_{NF}.
\]

The \emph{natural boundary condition} minimizes the loss function and leads to
an optimal choice of the policy instrument at the initial date. This first
order condition is equivalent to state that the Lagrange multiplier (co-state
variable) of forward-looking inflation $\gamma_{t}$ is a backward-looking
variable. It is predetermined at its optimal value equal to zero at the
initial date. Because the Lagrange multiplier of inflation is a function of
the policy instrument, this implies that the policy instrument is optimally
predetermined at the initial date at the value $x_{0}^{\ast}$:%

\[
\left(  \frac{\partial L}{\partial\pi_{t}}\right)  _{\pi=\pi_{0}}=\gamma
_{0}=\gamma_{0}^{\ast}=0\Rightarrow x_{0}^{\ast}=-\varepsilon\frac{\lambda
}{1-\beta\rho\lambda}z_{0}.
\]

The policy rule is used to anchor initial inflation $\pi_{0}$ on the policy
instrument and the initial value of the autoregressive shock:%

\[
\pi_{0}=F_{\pi}^{\ast-1}x_{0}^{\ast}-F_{\pi}^{\ast-1}F_{z}^{\ast}z_{0}%
=-\frac{1}{\varepsilon}x_{0}^{\ast}=\frac{\lambda}{1-\beta\rho\lambda}z_{0}.
\]

Expected impulse response functions are given in table 1. For given values of
the policy parameter $\alpha>0$ and of transmission parameters $\left(
\beta,\kappa\right)  $, Ramsey optimal policy has reduced form rule parameters
which are \emph{observationally equivalent} to the ones of a simple rule such
that $F_{\pi}=F_{\pi}^{\ast}$ $\in D_{NF}^{\ast}$, $F_{z}=F_{z}^{\ast}$, and a
backward-looking (predetermined) policy instrument such that its initial value
is $x_{0}=x_{0}^{\ast}$.

Let's now do a thought experiment assuming that inflation is a
backward-looking variable with known initial value $\overline{\pi}_{0}$ (in
mathematics, an "\emph{essential natural condition}"). The recursive dynamics
of the new-Keynesian Phillips curve can now be interpreted as
backward-looking, assuming expected inflation is caused by the lagged value of
inflation. The only change for optimal policy is that an initial boundary
condition is an essential boundary condition $\pi_{0}=\overline{\pi}_{0}$
instead of a natural boundary condition. This does not change the optimal
values of the policy rule parameters $\left(  F_{\pi}^{\ast},F_{z}^{\ast
}\right)  $ which do not depend on initial values $\left(  x_{0},\pi_{0}%
,z_{0}\right)  $. In this case, the optimal policy rule determines the initial
value of the policy instrument, knowing the initial value of the policy target:%

\[
x_{0}=F_{\pi}^{\ast}\overline{\pi}_{0}+F_{z}^{\ast}z_{0}.
\]

To sum up, if inflation is forward-looking, its Lagrange multiplier is
optimally predetermined. The optimal decision of $x_{0}^{\ast}$ according to
the natural boundary conditions anchors initial inflation $\pi_{0}$ using the
optimal policy feedback rule.

Conversely, if we do the thought experiment that inflation is
backward-looking, its policy maker's Lagrange multiplier is forward-looking.
The optimal initial value of the policy instrument $x_{0}$ is found using the
optimal policy feedback rule: it jumps as a forward-looking variable.

\subsection{Simple rule with forward-looking policy instrument}

In the next step, we assume that both, the policy target and the policy
instrument, are forward-looking variables without initial values. The
cost-push forcing variable is then the only predetermined variable. Blanchard
and Kahn's (1980) determinacy condition leads to the requirement that one
eigenvalue should lie strictly inside the unit circle and one eigenvalue
outside the unit circle for the closed-loop system. The eigenvalue of the cost
push shock is assumed to be inside the unit circle $0<\rho<1$. Then the policy
rule parameter $F_{\pi}$ for given transmission parameters $\kappa$ and
$\beta$, should be chosen such that the eigenvalue of the inflation dynamics
of the closed loop system is outside the unit circle: $\lambda_{SR}%
=\frac{1-\kappa F_{\pi}}{\beta}>1$ or $\frac{1-\kappa F_{\pi}}{\beta}<-1$.
These "positive-feedback" values of the policy-rule parameter enforces the
local instability of the closed-loop dynamical system of dimension two.

This implies the following "positive-feedback determinacy set" $D_{PF}$ which
is the complement of the determinacy set $%
%TCIMACRO{\U{211d} }%
%BeginExpansion
\mathbb{R}
%EndExpansion
\backslash D_{NF}=\overline{D_{NF}}$ with respect to the case when the policy
instrument is backward-looking.%

\begin{align*}
F_{\pi}  &  \in D_{PF}=\overline{D_{NF}}=\left\{  F_{\pi}\in%
%TCIMACRO{\U{211d} }%
%BeginExpansion
\mathbb{R}
%EndExpansion
\text{ such that }\left\vert \frac{1-\kappa F_{\pi}}{\beta}\right\vert
\geq1\text{ }\right\}  \Rightarrow\\
\overline{D_{NF}}  &  =\left]  -\infty,\frac{1-\beta}{\kappa}\right]
\cup\left[  \frac{1+\beta}{\kappa},+\infty\right[  \text{ .}%
\end{align*}

For given values of the transmission mechanism $\left(  \beta,\kappa\right)
$, the policy-rule parameter $F_{\pi}$ is a bifurcation parameter. Indeed, if
$\left(  \beta,\kappa\right)  $ are not given, they are also bifurcation
parameters of the closed-loop system. However, the common practice in
new-Keynesian models is to present the determinacy sets $D_{PF}$ for the
feedback-rule parameters (Woodford (2003)).

When $F_{\pi}=\frac{1-\beta}{\kappa}+\eta$, with $\eta$ a small real number,
shifts with a small change to $F_{\pi}=\frac{1-\beta}{\kappa}-\eta$, the
dynamical system shifts from having $\lambda=\frac{1-\kappa F_{\pi}}{\beta}<1$
to $\lambda\geq1$ which corresponds to a saddle-node bifurcation. When
$F_{\pi}=\frac{1+\beta}{\kappa}-\eta$, with $\eta$ a small real number, shifts
with a small change to $F_{\pi}=\frac{1+\beta}{\kappa}+\eta$, the dynamic
system shifts from having $\lambda>-1$ to $\lambda\leq-1$ which corresponds to
a flip bifurcation. A\ bifurcation is a qualitative change of the behavior of
the dynamical system.

Both, inflation and the policy instrument (output gap), have to be
proportional to the cost-push auto-regressive forcing variable. Then, for
identification the policy rule $F_{z}$ has to be restricted to zero, as it is
useless to respond to two variables when the dynamics evolves in a space of
dimension one. Using Blanchard and Kahn's (1980) method, the solution is given
by the slope of the eigenvectors of the given stable eigenvalue $0<\rho<1$ of
the cost-push shock:%

\begin{equation}
\left(
\begin{array}
[c]{cc}%
\frac{1-\kappa F_{\pi}}{\beta} & -\frac{1}{\beta}\\
0 & \rho
\end{array}
\right)  \left(
\begin{array}
[c]{c}%
\pi_{t}\\
z_{t}%
\end{array}
\right)  =\rho\left(
\begin{array}
[c]{c}%
\pi_{t}\\
z_{t}%
\end{array}
\right)  \Rightarrow\left(  \frac{1-\kappa F_{\pi}}{\beta}-\rho\right)
\pi_{t}=\frac{1}{\beta}z_{t}.
\end{equation}

The solution is such that:%

\[
\pi_{t}=\frac{\frac{1}{\beta}}{\frac{1-\kappa F_{\pi}}{\beta}-\rho}z_{t}%
=\frac{1}{1-\kappa F_{\pi}-\rho\beta}z_{t}\text{ and }x_{t}=\frac{\frac
{1}{\beta}F_{\pi}}{\frac{1-\kappa F_{\pi}}{\beta}-\rho}z_{t}\text{ for
}t=0,1,2,...
\]

The dynamical system is given in table 1 on the third row with $x_{0}$ unknown.

\subsection{Discretion}

If the policy maker re-optimizes with certainty in each future periods we call
this an infinite-horizon zero-credibility policy. This zero-credibility policy
is labeled "discretionary policy" (Clarida, Gali and Gertler (1999)) although
there are other definitions of discretionary policy (Chatelain and Ralf
(2020d)). It is equivalent to the optimal simple rule for the new-Keynesian
Phillips curve.

When the probability of not reneging commitment is exactly zero ($q=0$), the
policy maker's loss function boils down to a \emph{static utility} ($\left(
\beta.0\right)  ^{0}=1$, $\left(  \beta.0\right)  ^{t}=0,t=1,2,...$): the
policy maker \emph{only values the current instantaneous period}, as he knows
he will be replaced next period, whatever the duration of the next period:%

\[
V^{jk}\left(  u_{0}\right)  =-E_{0}%
%TCIMACRO{\dsum \limits_{t=0}^{t=+\infty}}%
%BeginExpansion
{\displaystyle\sum\limits_{t=0}^{t=+\infty}}
%EndExpansion
\left(  \beta q\right)  ^{t}\left[  \frac{1}{2}\left(  \pi_{t}^{2}%
+\frac{\kappa}{\varepsilon}x_{t}^{2}\right)  +\beta\left(  1-q\right)
V^{jk}\left(  u_{t}\right)  \right]  =-\frac{1}{2}\left(  \pi_{0}^{2}%
+\frac{\kappa}{\varepsilon}x_{0}^{2}\right)  .
\]

In the transmission mechanism, the policy maker knows for sure that another
policy maker will take his office. For him, the new-Keynesian Phillips curve
boils down to a \emph{static Phillips curve} where he sets a probability zero
of taking into account expected inflation:%

\[
\text{s.t. }\pi_{0}=\kappa x_{0}+v_{0}+\left(  \beta.0\right)  E_{t}\pi
_{1}\text{ where }v_{0}=\beta\left(  1-0\right)  E_{0}\left[  \pi_{1}%
^{k}\right]  +u_{0}~\text{with }u_{0}\text{ given.}%
\]

Gali (2015) interprets this term as follows: "\textit{the term }$v_{t}=\beta
E_{t}\left[  \pi_{t+1}\right]  +z_{t}$\textit{ is taken as given by the
monetary authority, because }$z_{t}$\textit{ is exogenous and }$E_{t}\left[
\pi_{t+1}^{k}\right]  $\textit{ is a function of expectations about future
output gaps (as well as future }$z_{t}$\textit{'s) which, by assumption,
cannot be currently influenced by the policymaker". }

As expected inflation is not taken into account by the policy maker $j$,
\emph{this removes one order in the dynamics of inflation} in the transmission
mechanism, which removes one dimension of the dynamical system. Shifting from
a \emph{dynamic} new-Keynesian Phillips curve to a \emph{static} Phillips
curve is \emph{the origin of the bifurcation} of the dynamical system between
the case of extreme discretion ($q=0$) and the alternative case of
quasi-commitment ($q\in\left]  0,1\right]  $), where at least a non-zero
weight is set on expected inflation by the policy maker.

The optimality condition implies a policy rule with perfect negative
correlation of the policy instrument (output gap) with the policy target
(inflation) with a constant parameter given by the opposite of the household's
elasticity of substitution between goods:%

\[
x_{t}=F_{\pi}\pi_{t}\text{ with }F_{\pi}=-\frac{\kappa}{\frac{\kappa
}{\varepsilon}}=-\varepsilon\text{ }\in D_{PF}^{\ast}=\left]  -\infty
,-1\right[  \text{ because }\varepsilon\in\left]  1,+\infty\right[  \text{,
for }t=0,1,2,...
\]

Clarida,\ Gali and Gertler (1999) and Gali (2015), chapter 5, assume that
both, the policy instrument and the policy target, are forward looking and
that the cost-push shock is the only predetermined variable. Then Blanchard
and Kahn's (1980) determinacy solution is given by the unique slope of the
eigenvectors of the given stable eigenvalue $0<\rho<1$ of the cost-push shock:%

\begin{equation}
\left(
\begin{array}
[c]{cc}%
\frac{1}{\beta}+\frac{\kappa}{\beta}\varepsilon & -\frac{1}{\beta}\\
0 & \rho
\end{array}
\right)  \left(
\begin{array}
[c]{c}%
\pi_{t}\\
z_{t}%
\end{array}
\right)  =\rho\left(
\begin{array}
[c]{c}%
\pi_{t}\\
z_{t}%
\end{array}
\right)  \Rightarrow\left(  \frac{1}{\beta}+\frac{\kappa}{\beta}%
\varepsilon-\rho\right)  \pi_{t}=\frac{1}{\beta}z_{t}.
\end{equation}

The discretion solution for the policy target and the policy instrument are
both proportional to the only predetermined variable: the cost-push
auto-regressive shock:%

\begin{equation}
\pi_{t}=\left(  \frac{1}{1-\beta\rho+\kappa\varepsilon}\right)  z_{t}\text{
and }x_{t}=-\varepsilon\left(  \frac{1}{1-\beta\rho+\kappa\varepsilon}\right)
z_{t}\text{. }%
\end{equation}

When varying the policy maker's preferences ($\varepsilon\in\left]
1,+\infty\right[  $), a subset $D_{PF}^{\ast}$ of the positive-feedback
determinacy set $D_{PF}$ of the policy-rule parameters is the same as for the
simple rule, assuming that both, the policy target and the policy instrument,
are forward-looking which corresponds to a reduced form of discretion:%

\[
F_{\pi}\in D_{PF}^{\ast}=\left]  -\infty,-1\right[  \subset D_{PF}%
=\overline{D_{NF}}=\left]  -\infty,\frac{1-\beta}{\kappa}\right]  \cup\left[
\frac{1+\beta}{\kappa},+\infty\right[  \text{ }%
\]

The assumption of forward-looking policy instrument for forward-looking policy
target corresponds to a policy maker's static model \emph{where the weight on
the private sector rational expectations is zero}. When setting his policy
instrument, the central banker does not care at all about the future.

\subsection{Bifurcation and robustness}

For all non-zero probabilities of the policy maker to remain in office,
$\left\{  q\in\left]  0,1\right]  \right\}  $, the monetary policy corresponds
to Ramsey optimal policy under quasi-commitment. Even an extremely small level
of credibility (e.g. a probability of not reneging commitment equal to
$10^{-7}$) still involves taking into account the expectations of inflation by
the policy maker. This means taking into account an additional dimension of
the dynamical system (order two) as compared to extreme discretion. Only the
event $\left\{  q=0\right\}  $ corresponds to a Dirac distribution, with a
zero prior probability for this event.

With quasi-commitment, the probability of not reneging commitment can be
infinitely small (near-zero credibility), but it remains strictly positive:
for example, $q=10^{-7}>0$ with $q\in\left]  0,1\right]  $, hence $\beta
q\in\left]  0,0.99\right]  $. The following table 1 compares the solutions of
the different scenarios of backward or forward looking instruments and targets:

\textbf{Table 1:} Expected impulse response functions: Backward-looking policy
instrument, Ramsey optimal policy, forward-looking policy instrument.

\qquad%
\begin{tabular}
[c]{|l|l|l|}\hline
& Expected impulse response functions following $z_{0}$ & $\frac{1-\kappa
F_{\pi}}{\beta}$\\\hline
$x_{0}$ given & $\left(
\begin{array}
[c]{c}%
\pi_{t}\\
z_{t}%
\end{array}
\right)  =\left(
\begin{array}
[c]{cc}%
\frac{1-\kappa F_{\pi}}{\beta}=\lambda & -\frac{1}{\beta}-\frac{\kappa}{\beta
}F_{z}\\
0 & \rho
\end{array}
\right)  ^{t}\left(
\begin{array}
[c]{c}%
-F_{\pi}^{-1}F_{z}z_{0}-F_{\pi}^{-1}x_{0}\\
z_{0}%
\end{array}
\right)  $ & $\left\vert \lambda_{IBL}\right\vert <1$\\\hline
$x_{0}^{\ast}=-\varepsilon\frac{\lambda}{1-\beta\rho\lambda}z_{0}\ \ $ &
$\left(
\begin{array}
[c]{c}%
\pi_{t}\\
z_{t}%
\end{array}
\right)  =\left(
\begin{array}
[c]{cc}%
\frac{1-\kappa F_{\pi}}{\beta}=\lambda & -\frac{1}{\beta}-\frac{\kappa}{\beta
}F_{z}\\
0 & \rho
\end{array}
\right)  ^{t}\left(
\begin{array}
[c]{c}%
\frac{\lambda}{1-\beta\rho\lambda}z_{0}\\
z_{0}%
\end{array}
\right)  $ & $0<\lambda<1$\\\hline
$x_{0}$ unknown & $\left(
\begin{array}
[c]{c}%
\pi_{t}\\
z_{t}%
\end{array}
\right)  =\left(
\begin{array}
[c]{cc}%
\frac{1-\kappa F_{\pi}}{\beta}=\lambda_{IFL} & -\frac{1}{\beta}\\
0 & \rho
\end{array}
\right)  ^{t}\left(
\begin{array}
[c]{c}%
\frac{1/\beta}{\lambda_{IFL}-\rho}z_{0}\\
z_{0}%
\end{array}
\right)  $ & $\left\vert \lambda_{IFL}\right\vert >1$\\\hline
Discretion & $\left(
\begin{array}
[c]{c}%
\pi_{t}\\
z_{t}%
\end{array}
\right)  =\left(
\begin{array}
[c]{cc}%
\frac{1-\kappa F_{\pi}}{\beta}=\lambda_{D} & -\frac{1}{\beta}\\
0 & \rho
\end{array}
\right)  ^{t}\left(
\begin{array}
[c]{c}%
\frac{1/\beta}{\lambda_{IFL}-\rho}z_{0}\\
z_{0}%
\end{array}
\right)  $ & $\left\vert \lambda_{D}\right\vert >1$\\\hline
\end{tabular}

Table 2 shows numerical examples for the different scenarios.

\textbf{Table 2:} Expected impulse response functions for $\rho=0.8$,
$\beta=0.99$, $\varepsilon=6$, $\kappa=0.1275$ obtained with $\theta=2/3$,
$1-\alpha_{L}=2/3$, $\sigma=1$ and $\varphi=1$.%

\begin{tabular}
[c]{|l|l|l|l|}\hline
& Expected impulse response functions following $z_{0}$ & $\frac{1-\kappa
F_{\pi}}{\beta}$ & $F_{\pi}\in$\\\hline
$x_{0}$ $=0.65z_{0}$\ given & $\left(
\begin{array}
[c]{c}%
\pi_{t}\\
z_{t}%
\end{array}
\right)  =\left(
\begin{array}
[c]{cc}%
0.43 & -0.13\\
0 & 0.8
\end{array}
\right)  ^{t}\left(
\begin{array}
[c]{c}%
-F_{\pi}^{-1}F_{z}z_{0}-F_{\pi}^{-1}x_{0}\\
z_{0}%
\end{array}
\right)  $ & $0.43<1$ & $4.51\in D_{NF}$\\\hline
$x_{0}^{\ast}=-\varepsilon\frac{\lambda}{1-\beta\rho\lambda}z_{0}\ \ $ &
$\left(
\begin{array}
[c]{c}%
\pi_{t}\\
u_{t}%
\end{array}
\right)  =\left(
\begin{array}
[c]{cc}%
0.43 & -0.13\\
0 & 0.8
\end{array}
\right)  ^{t}\left(
\begin{array}
[c]{c}%
0.65\\
1
\end{array}
\right)  z_{0}$ & $0.43<1$ & $4.51\in D_{NF}$\\\hline
$x_{0}$ unknown & $\left(
\begin{array}
[c]{c}%
\pi_{t}\\
u_{t}%
\end{array}
\right)  =\left(
\begin{array}
[c]{cc}%
1.78 & -1.01\\
0 & 0.8
\end{array}
\right)  ^{t}\left(
\begin{array}
[c]{c}%
1.03\\
1
\end{array}
\right)  z_{0}$ & $1.78>1$ & $-6\in\overline{D_{NF}}$\\\hline
Discretion & $\left(
\begin{array}
[c]{c}%
\pi_{t}\\
u_{t}%
\end{array}
\right)  =\left(
\begin{array}
[c]{cc}%
1.78 & -1.01\\
0 & 0.8
\end{array}
\right)  ^{t}\left(
\begin{array}
[c]{c}%
1.03\\
1
\end{array}
\right)  z_{0}$ & $1.78>1$ & $-6\notin\overline{D_{NF}}$\\\hline
\end{tabular}

Giordani and S\"{o}derlind (2006) demonstrated that the simple-rule solution
with forward-looking policy instruments and forward-looking policy targets
does not satisfy robustness to misspecification as defined by Hansen and
Sargent (2008). The evil agent of robust control trying to fool the policy
maker as much as he can is turned into the \emph{most helpful angel agent}
which tries to help as much as he can the policy maker to save his locally
unstable path in the space of policy target variables. Macroeconomists advise
policy makers that the local instability of the dynamics of the policy maker's
state variables is necessary in order to achieve macroeconomic stabilization.
This solution is turning upside down the theories of classic, optimal and
robust control and of Stackelberg dynamic games. Its policy recommendations
have zero robustness and zero credibility (Cochrane (2019) p.345).

The uncertainty on the knowledge of parameters is strictly restricted, so that
the local instability of the equilibrium in the space of $n$ state variables
of the solution with forward-looking hypothesis does not matter. A\ shock on
one parameter is instantaneously offset by another shock on another parameter
(by the virtue of the angel agent) so that the dynamics of the $n$ state
variables remains in a stable subset of dimension $m<n$\emph{. }

In our example, assume that the policy rule $F_{\pi}$ is chosen by the policy
maker and the stable subspace projection parameter $G$ anchoring inflation on
the non-observable cost-push auto-regressive shock is chosen by the private
sector.%
\[
\pi_{t}=Gz_{t}\text{ and }x_{t}=F_{\pi}\pi_{t}\text{ with }G=\frac{1}{1-\kappa
F_{\pi}-\rho\beta}.
\]

This solution remains stable only if the uncertainty on the parameters $\beta
$,$\kappa$ and $\rho$ satisfies this restriction:%

\[
\kappa=\frac{1}{F_{\pi}}\left(  \frac{1}{G}+\rho\beta-1\right)  .
\]

A deviation of $\kappa$ should be instantaneously compensated by a deviation
of $\rho\beta$ so that $G$ and $F_{\pi}$ remain unchanged. \ \emph{There is
the underlying assumption of a angel agent who creates an exact
negative-feedback correlation between exogenous shocks. Hence, the local
instability in the space of policy targets of dimension two does not change
the stable dynamics of inflation and of the policy instrument in a subspace of
dimension one. }When $\kappa$, $\beta$, $\rho$ vary in continuous intervals of
real numbers, this restriction has a probability equal to zero. Robustness of
misspecification can also be ruled out by assuming the knowledge of all the
parameters of the policy transmission mechanism with infinite precision.

\section{Conclusion}

After having analyzed the different scenarios of forward or backward looking
policy instruments and their relation to the policy targets, we will propose
two criteria for choosing the appropriate economic model. The first concerns
the robustness of the model:

\begin{criterion}
A model of policy maker's behavior should provide policy recommendations which
are robust to misspecification when the policy maker is facing uncertainty on
the values of the parameters of the economy. A necessary (but not sufficient)
condition is to use negative-feedback policy rule parameters.
\end{criterion}

The second criterion concerns the way the central bank's behavior is modeled.
Simple rules with proportional feedback as used in New-Keynesian DSCG\ models
where a policy instrument responds to forward-looking policy targets does not
imply that policy instruments are forward-looking variables. These models are
sometimes considered to have better theoretical foundations due to optimal
behavior with respect to agent-based models using behavioral simple rules.
Consistency implies that this argument of the rational micro-foundation of
optimal choice should also apply to the policy maker. The first best method is
to abandon once and for all ambiguous simple-rule models and to shift to
models of Ramsey optimal policy under quasi-commitment. If simple rules are
still assumed as a second best theory, the outcome of their model should be as
close as possible to Ramsey optimal policy under quasi-commitment.

\begin{criterion}
Simple-rule parameters should be reduced forms of the rules of Ramsey optimal
policy under quasi-commitment, which would imply a minimal robustness to
misspecification and a minimal credibility of stabilization policy.
\end{criterion}

As seen in the example, this criterion can be satisfied with the
backward-looking hypothesis for a range of values of the policy-rule parameters.

This paper suggest that new-Keynesian DSGE models and the fiscal theory of the
price level\ are currently \emph{locked in} to the inferior technology path of
having chosen the convention of forward-looking policy instrument if policy
targets are forward-looking, as it happened for QWERTY\ typewriter keyboard
with respect to DVORAK (David (1985)).

\end{document}